\documentclass[runningheads]{llncs}

\usepackage{apacite}

\usepackage[utf8]{inputenc}

\usepackage[T1,T2A]{fontenc}
\usepackage[table]{xcolor}
\usepackage{colortbl}
\usepackage[ukrainian,english]{babel}
\usepackage{url}
\usepackage{listings}

\usepackage{float} 

\usepackage{graphicx}
\begin{document}

\title{The OpenCitations Index}
%
%
\author{Ivan Heibi\inst{1,2}\orcidID{0000-0001-5366-5194} \and
Arianna Moretti\inst{1}\orcidID{0000-0001-5486-7070} \and \\
Silvio Peroni\inst{1,2}\orcidID{0000-0003-0530-4305} \and
Marta Soricetti\inst{1}\orcidID{0009-0008-1466-7742}}
\authorrunning{I. Heibi et al.}
%
\institute{Research Centre for Open Scholarly Metadata\and Digital Humanities Advanced Research Centre (DHARC)\\Department of Classical Philology and Italian Studies, University of Bologna, via Zamboni 32, 40126, Bologna, BO, Italy\\
\email{\{ivan.heibi2, arianna.moretti4, silvio.peroni, marta.soricetti2\}@unibo.it}
}
\maketitle   

%
\begin{abstract}
This article presents the OpenCitations Index, a collection of open citation data maintained by OpenCitations, an independent, not-for-profit infrastructure organisation for open scholarship dedicated to publishing open bibliographic and citation data using Semantic Web and Linked Open Data technologies. The collection involves citation data harvested from multiple sources. To address the possibility of different sources providing citation data for bibliographic entities represented with different identifiers, therefore potentially representing same citation, a deduplication mechanism has been implemented. This ensures that citations integrated into OpenCitations Index are accurately identified uniquely, even when different identifiers are used. This mechanism follows a specific workflow, which encompasses a preprocessing of the original source data, a management of the provided bibliographic metadata, and the generation of new citation data to be integrated into the OpenCitations Index. The process relies on another data collection – OpenCitations Meta, and on the use of a new globally persistent identifier, namely OMID (OpenCitations Meta Identifier). As of July 2024, OpenCitations Index stores over 2 billion unique citation links, harvest from Crossref, the National Institute of Heath Open Citation Collection (NIH-OCC), DataCite, OpenAIRE, and the Japan Link Center (JaLC). OpenCitations Index can be systematically accessed and queried through several services, including SPARQL endpoint, REST APIs, and web interfaces. Additionally, dataset dumps are available for free download and reuse (under CC0 waiver) in various formats (CSV, N-Triples, and Scholix), including provenance and change tracking information.

\end{abstract}
\section{Introduction}
Bibliographic citations are conceptual links from a citing to a cited entity. Their relevance lies in depicting global research endeavours, unveiling trends in scholarly knowledge across time, assigning credit to authors, and evaluating their impact \cite{peroni_open_2018}. Recently, to provide open and transparent alternatives to well-known proprietary services, several initiatives and infrastructures have started to collect citation information and publish them under permissive licenses and in machine-readable formats -- a necessary condition to foster reproducibility, ethical research conduct, and fair engagement in academic knowledge \cite{sugimoto_open_2017}. OpenCitations (\url{https://opencitations.net}) \cite{peroni_opencitations_2020} is one of the infrastructures that has been created to pursue that purpose.

One of the first collections OpenCitations provided, back in 2018, was the OpenCitations Index of Crossref open DOI-to-DOI citations (COCI) \cite{heibi_software_2019}. This collection of citations contained DOI-to-DOI citation links (i.e. \emph{entity A cites entity B}) gathered from articles' reference lists included in Crossref reference lists \cite{hendricks_crossref_2020}. The release of COCI marked a significant milestone for the infrastructure and the Open Science community as a whole, since it offered a large machine-readable collection of open citation data alternative to popular subscription-oriented sources. 

All citations in COCI  were defined according to the OpenCitations Data Model (OCDM) \cite{daquino2020opencitations} by means of Semantic Web technologies. According to the OCDM, citations are modelled as first-class entities having their own metadata -- citing entity, cited entity, citation creation date, citation timespan, etc. -- and provenance information -- the source of the original raw data, the agent responsible for the ingestion into OpenCitations, and the date for such activity). 



In the last few years, primary work has been dedicated to extending the coverage of open citation data available in OpenCitations collections by ingesting further information from new data sources (in addition to Crossref). In particular, from December 2022 to November 2023, we have re-engineered OpenCitations ingestion workflow and started to include citations coming from the National Institute of Health Open Citation Collection (NIH-OCC) \cite{hutchins_nih_2019}, which mainly contains mainly citations involving PubMed articles, DataCite, OpenAIRE and the Japan Link Center (JaLC) \cite{kato_gestione_2012}, where we started to work actively on the bibliodiversity availability of citation data by working on Japanese research publications \cite{moretti_integration_2024}. We included all citation data we ingested from these five distinct data sources in a new comprehensive collection named OpenCitations Index (\url{https://opencitations.net/index}).


This article aims to be the canonical reference for the OpenCitations Index. Here we outline this new collection and describe the methodological workflow enabling efficient and scalable ingestion of citation data from external sources. The creation of the OpenCitations Index is also described considering the main design choices that characterise the creation of OpenCitations Meta (\url{https://opencitations.net/meta}) \cite{massari_opencitations_2024}, which is another OpenCitations collection which stores and delivers bibliographic metadata for all the entities involved in the citations included in the OpenCitations Index.


The rest of the article is structured as follows. In Section~\ref{sec:related}, initiatives for disseminating citation data are illustrated, focusing on semantic datasets. Section~\ref{sec:source} introduces and discusses the external sources used to harvest citation data to be stored in OpenCitations. Then, the OpenCitations Index is introduced in Section~\ref{sec:index}, where we discuss the reasons for creating a single index of unique citations and explain how this is planned to be done. We then present the dataset and its ancillary collections, i.e., citation collection and provenance information, along with the resources and services used to make the data of the OpenCitations Index available. Section~\ref{sec:usage} provides evidence of utilising the data produced through quantitative and qualitative analysis, with particular attention to community uptake. Finally, in Section~\ref{sec:conclusions}, we present final remarks and future directions.

\section{Related Works}\label{sec:related}

In recent years, the evolution of scholarly communication platforms and databases reflects ongoing efforts to increase the open accessibility and interconnectivity of scholarly works, particularly after the launch of the Initiative for Open Citations (I4OC) (\url{https://i4oc.org/}) in 2017, followed by the decision by Crossref, in 2022, to require all publishing depositing bibliographic references making them openly accessible \cite{hendricks_amendments_nodate}. This scenario, coupled with all the other Open Science initiatives around research information and bibliographic metadata, has created and launched several infrastructures responsible for hosting and making available open bibliographic and citation data.

The \textit{Open Ukrainian Citation Index} (OUCI, \url{https://ouci.dntb.gov.ua/en/}) manages information regarding around 157 million international publications and more than 587,191 publications in 1,848 Ukrainian journals (as of 7 May 2024). The primary aim of OUCI is enhancing access to scholarly citation data by exploiting DOIs to improve the state of the art in bibliometrics studies, especially at a national level \cite{__2023}.

OpenAIRE (\url{https://openaire.eu}), is a non-profit organisation dedicated to promoting open scholarship and enhancing data discoverability, accessibility, shareability, reusability, reproducibility \cite{manghi2010infrastructure}.  OpenAIRE gathers bibliographic citation metadata from various sources (such as libraries, publication repositories, publishers, and author directories) to populate a research graph storing scientific products (e.g., results, organizations,data sources, etc) \cite{manghi2019openaire}.  As of July 2024, OpenAIRE reported 272.6 million research products in its graph.


Created from the ashes of Microsoft Academic Graph, dismissed by Microsoft in 2021, \textit{OpenAlex} (\url{https://openalex.org/})  \cite{priem_openalex_2022} is a catalogue that aims to provide a reliable, free, and open data source for bibliographic and citation data. The index includes over 250 million works linked by over a billion connections, accessible via a web interface, API, or database snapshot. The catalogue includes a description of works, authors, venues, institutions, and concepts woven together in a directed graph.

In medical and biomedical literature, several national and international institutions have put a lot of effort into providing reliable bibliographic metadata describing scholarly publications. For instance, \textit{iCite} (\url{https://icite.od.nih.gov/}) is a web-based tool provided by the National Institute of Health to allow users to access a database of citation data for publications identified by PMIDs \cite{hutchins_nih_2019}. As of March 2024, the project’s dataset contains 31,08 million publications from the 1980s onward.


Within the introduction of Wikidata \cite{vrandecic_wikidata_2014}, the community working on bibliographic information grouped together in an internal initiative named \textit{WikiCite} (\url{http://wikicite.org/}). This initiative exploits Wikidata to enhance, augment, and refine citations in Wikimedia Foundation's projects, particularly Wikipedia. The principles guiding WikiCite are the openness of data and the will to create a collaborative environment to support the growth of the bibliometrics study field by developing innovative tools. The most updated data officially exposed by WikiCite dates back to the 26th of December 2022 and declares more than 287 million bibliographic entities covered, while active work on the initiative is still maintained by the community, e.g. Scholia \cite{nielsen_scholia_2017}.

Finally, \textit{Semantic Scholar} (\url{https://www.semanticscholar.org/}) \cite{kinney_semantic_2023} was launched in 2015 as a free research service for scientific literature based on artificial intelligence (AI), and currently contains more than 214 million papers, covering most scientific fields. The project exploits AI to improve literature discovery, analysis, and understanding by enhancing search accuracy through smart search. It provides personalized recommendations based on user behaviour and enhances the analysis of citations for assessing a paper's impact, conducting semantic analysis to identify research trends and connections. Access to data is guaranteed via the website, the Semantic Scholar API, and the Open Research Corpus.


\section{Authoritative citation data sources}\label{sec:source}
The citation data managed, ingested, and maintained by OpenCitations have all been harvested from external data sources. Each source provides its raw citation data under a CC0 waiver or a similar license, which is mandatory request in order to be compliant with the policies of OpenCitations. This has been the case since the inception of OpenCitations, with the first collections released in the past years \cite{damato_one_2017}. We briefly introduce the five sources currently used to create the OpenCitations Index, some of which were introduced in the previous section (~\ref{sec:related}). This aligns with the OpenCitations' guiding principles, which actively promote cooperation with similar sources to advance the principles of Open Science. By collaborating with other initiatives, OpenCitations aims to provide a better service and support the broader open science community.

\subsubsection{Crossref} 
Crossref (\url{https://crossref.org}) \cite{hendricks_crossref_2020} was the first data source from which OpenCitations created an index of citation data expressed as first-class entities with their own metadata, namely COCI \cite{heibi_software_2019}, and currently is the major provider of citation data included in the OpenCitations Index. Crossref is a DOI registration agency and a global community infrastructure, counting up more than 158 million records (as of 12 May 2024). The persistent identifiers adopted for the bibliographic entities are DOIs, the records accessible via the Crossref API are represented as JSON documents and citation data are provided as metadata for the citing entities. OpenCitations downloads and processes Crossref data, made available as dumps, periodically -- usually every two months, unless specific issues occur.


\subsubsection{NIH Open Citation Collection}
The National Institute of Health Open Citation Collection (NIH-OCC) stores bibliographic and citation data from the biomedical domain. The data are published by the Office for Portfolio Analysis of the Office of the Director of the U.S. National Institutes of Health and provided periodically on FigShare, an open-access repository for research material \cite{hutchins_nih_2019}. The NIH-OCC mainly exposes citations between articles indexed in PubMed, which can also be retrieved by the iCite web service (\url{https://icite.od.nih.gov}) \cite{icite_icite_2022}. The dataset consists of two subsets: the Open Citation Collection itself and the iCite Metadata. While the Open Citation Collection only exposes citation links, the iCite Metadata stores metadata of bibliographic resources (e.g., articles), including citation information.

Unlike Crossref, which uses DOIs, the persistent identifier used by NIH-OCC to refer to the bibliographic resources it contains is the PubMed Identifier (PMID). 
As a pitfall, the dataset does not adopt specific persistent identifiers for other bibliographic entities, such as authors and journals -- the latter are represented by using only their abbreviated names. In addition, other relevant information concerning the bibliographic resources in the collection is missing, such as the publisher's name, the editor, the page, volume and issue numbers, and the complete publication date beyond the year. This is partially justifiable since PubMed, the original source used to create NIH-OCC, is not a journal per se but an index of biomedical publications gathered from the related journals where they have been officially published. Thus, additional metadata about the journals involved are external to NIH-OCC data and, if needed, should be retrieved using external services. 

\subsubsection{DataCite}
DataCite (\url{https://datacite.org}) is a DOI registration agency created originally to provide persistent identifiers for research data. It makes a REST API available \cite{fenner_common_2016}, enabling users to ask for bibliographic and citation data while they do not officially release dumps. However, a dataset of bibliographic metadata based on DataCite is shared by the Internet Archive team on their portal once a year (e.g. at \url{https://archive.org/details/datacite-2024-01-26}). This dataset is serialized as JSON files using a new line delimiter (NLD-JSON), such that each line represents a bibliographic resource, according to the DataCite metadata schema \cite{datacite_metadata_working_group_datacite_2024}. 
Differently from the previous two sources, in DataCite citation links can be expressed either in the citing entities or the cited entities since the DataCite schema makes available direct (e.g. \emph{cites}) and inverse relations (e.g. \emph{isCitedBy}). The bibliographic resources included in DataCite are mainly identified by DOIs. Moreover, we can find additional identifiers in DataCite data, including ORCIDs for human agents, ISSNs for journals, and ISBNs for books.

\subsubsection{OpenAIRE ScholeXplorer}
OpenAIRE (\url{https://openaire.eu}) \cite{manghi_openaireplus_2012} is an European e-Infrastructure that promotes open access to research outputs and gathers bibliographic and other scholarly data from multiple sources. It offers query services to access the data systematically and periodically publishes dumps of its OpenAIRE graph. Part of the information contained in this graph come from ScholeXplorer (\url{https://scholexplorer.openaire.eu})  \cite{la_bruzzo_openaire_2022}, which is a service containing links (of different kinds) between traditional publications and datasets. These links are made available via a REST API, and a portion of these data, i.e. those being citation links, have also been published in Zenodo\cite{la_bruzzo_scholix_2023} in Scholix format  \cite{burton_scholix_2017} as a result of a collaboration between OpenAIRE and OpenCitations in the context of the EU-funded project OpenAIRE-Nexus (\url{https://doi.org/10.3030/101017452}).
The bibliographic resources included in the ScholeXplorer dump may specify multiple persistent identifiers (i.e. DOI, PMC, PMID, ArXiv, Handle). 


\subsubsection{Japan Link Center}
The Japan Link Center (JaLC, \url{https://japanlinkcenter.org}) is a DOI registration agency led by the Japan Science and Technology Agency (JST), aimed to manage in a centralized way bibliographic and citation information about digital academic contents provided by national institutions \cite{kato_gestione_2012}. It makes available a REST API for accessing its content and includes several metadata values in Japanese and English. 
All the bibliographic resources managed by JaLC are identified by DOIs. At the same time, they use a particular identifier schema, namely the J-STAGE identifier (JID), to refer to J-STAGE published journals (\url{https://www.jstage.jst.go.jp/}).

\section{The OpenCitations Index}
\label{sec:index}
In the previous section, we introduced the sources used to collect the citation data to be stored in OpenCitations. Here, we first define how citation data are modelled and then discuss the workflow that enables the ingestion of these data into OpenCitations. The workflow is designed to facilitate the transition of OpenCitations into a single index, which can be created and updated with all the unique citations harvested from multiple sources. Along with the actual citation data managed and stored, provenance/change tracking and descriptive metadata are produced, which are discussed in a dedicated subsection. The data gathered so far and the various services we offer for reusing and interacting with these citation data will be discussed in the latter parts of this section.

\subsection{Representing citation data}\label{sec:model}
Considering the need to manage large datasets of citation information from various, often overlapping sources, we developed specialised software components. These components facilitate metadata crosswalks between existing and potential new sources and the OpenCitations Data Model (OCDM), as well as the deduplication of entities to prevent multiple representations of the same citation link. For instance, the same citation between bibliographic resources may appear in multiple sources using different identifiers for such bibliographic resources. Therefore, we designed an extendable workflow to establish a globally unique procedure. This procedure is sufficiently customisable to capture the necessary information from each source, despite the possibility of having incompatible input formats and data models.

\begin{figure}
\includegraphics[width=\textwidth]{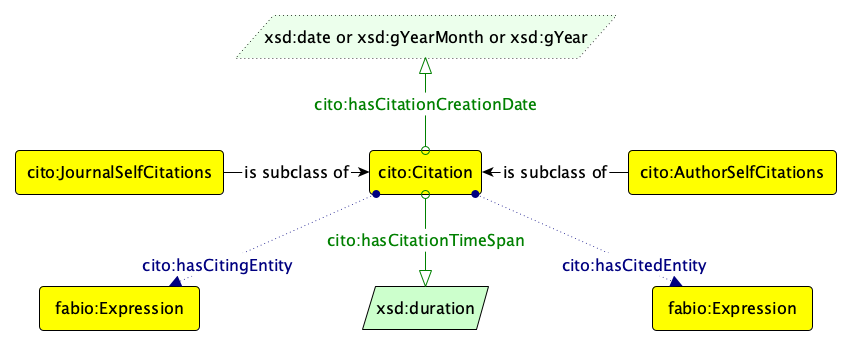}
\caption{The section of the OCDM that is dedicated to the description of citations. It details citations as first-class data entities accompanied by metadata, including the citing and cited entity, the citation creation date, the citation timespan, and self-citation information.}\label{ocdm_citations}
\end{figure}

This workflow aims to develop a process for ingesting new data into the OpenCitations Index, following the section of the OCDM dedicated to citation modelling; graphically represented in Figure~\ref{ocdm_citations}. The OCDM has been implemented via an OWL ontology, i.e. the OpenCitations Ontology (OCO, \url{https://w3id.org/oc/ontology}), that reuses several concepts defined in the \emph{Citation Typing Ontology} (CiTO, \url{http://purl.org/spar/cito}) \cite{peroni_fabio_2012}, part of the SPAR Ontologies (\url{http://www.sparontologies.net}) \cite{peroni_spar_2018}. In particular, the class \texttt{cito:Citation} is used to define a permanent conceptual directional link from the citing bibliographic entity to a cited bibliographic entity (i.e. a citation link: \emph{Entity\textsubscript{A} cites Entity\textsubscript{B}}).

The class \texttt{cito:Citation} enables a machine-readable specification for each citation, representing it as a first-class data entity with its own specified metadata, which includes:
\begin{itemize}
    \item the \emph{citing entity}, represented via the property \texttt{cito:hasCitingEntity}), that is the bibliographic resource which acts as the source for the citation (i.e. \emph{Entity\textsubscript{A}} in the citation link \emph{Entity\textsubscript{A} cites Entity\textsubscript{B}});
    \item the \emph{cited entity} (property \texttt{cito:hasCitedEntity}), i.e. the bibliographic resource target of the citation (i.e. \emph{Entity\textsubscript{B}} in the citation link \emph{Entity\textsubscript{A} cites Entity\textsubscript{B}});
    \item the \emph{citation creation date} (property \texttt{cito:hasCitationCreationDate}), that is the date on which the citation was created, correspondent to the publication date of \emph{Entity\textsubscript{A}};
    \item the citation timespan (property \texttt{cito:hasCitationTimespan}), i.e. the interval between \emph{Entity\textsubscript{A}}'s publication date and \emph{Entity\textsubscript{B}}'s publication date, represented using the XML Schema datatype for time duration (\url{https://www.w3.org/TR/xmlschema-2/#duration}).
    \item the \emph{type} of the citation, represented via two subclasses of \texttt{cito:Citation}, i.e. \texttt{cito:AuthorSelfCitation} and \texttt{cito:JournalSelfCitation}, that permits characterising when \emph{Entity\textsubscript{A}} and \emph{Entity\textsubscript{B}} have at least one author in common and when are published in the same journal, respectively.
\end{itemize}

\subsection{The ingestion workflow}
\begin{figure}
\includegraphics[width=\textwidth]{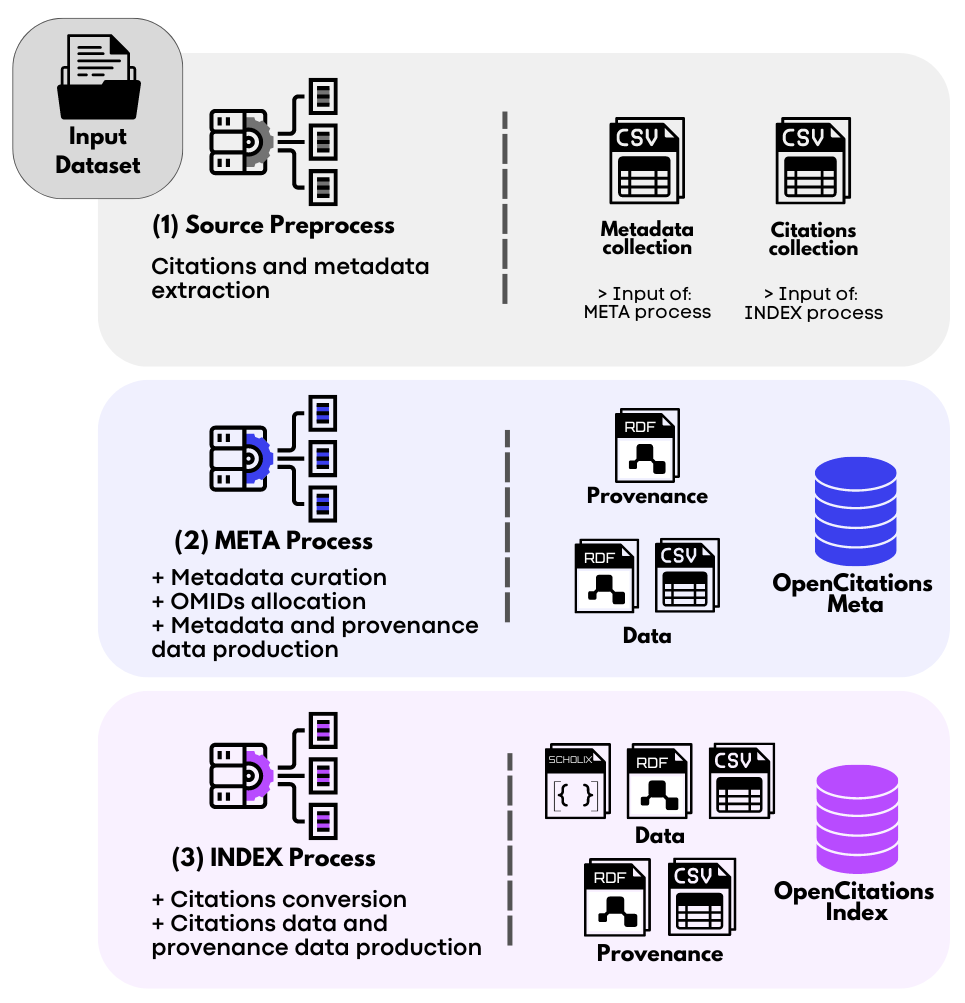}
\caption{The ingestion workflow of the OpenCitations Index; first, data provided by a specific source is processed to extract citations and metadata ("Source Preprocess"). This processed data is then fed into two sequential processes: the "META process" and the "INDEX process". The workflow leads to the dump production and online update of the OpenCitations Index dataset(triplestore).}\label{fig1}
\end{figure}

The ingestion workflow we developed aims to transform the data initially provided from the original source into to citation data formed following the data representation discussed in the previous section. The workflow is graphically represented in Figure~\ref{fig1}. Is based on three steps carried out sequentially: 

\begin{enumerate}
  \item \textit{Source Preprocess}, which extracts data from the original data source and produces tables in comma-separated value (CSV) format that will be used as input of the next steps;
  \item \textit{Meta Process}, which manages the metadata of the bibliographic resources involved in the citations;
  \item \textit{Index Process}, which generates the new citation data that will be included in the OpenCitations Index.
\end{enumerate}

The software implementing the workflow has been published with an open-source license on GitHub (\url{https://github.com/opencitations}) to foster the reproducibility of the ingestion. The following subsections detail each step by providing also a reference to the specific software component addressing the step.

The remainder of this section first addresses the preliminary step, followed by detailed explanations of the subsequent two macro processes in three separate sections. To foster the reuse of the workflow, all the software is open, reusable, and accessible via Github at: \url{https://github.com/opencitations}.

\subsubsection{Source Preprocess}
\label{sec:preliminaries}

This step is handled by the OpenCitations Data Sources Converter software \cite{oc_ds_converter_2024}. Here the aim is to provide a metadata crosswalk from the original source data model to the OCDM, validating identifiers and generating tabular information of bibliographic metadata and citation data to serve as input for the following two steps (i.e. META/INDEX process). 

The intrinsic complexity arising from the sources' diverse information content and organisational structures and the potential handling of novel identifiers, underscores the imperative for developing and using extensible software to fulfil possible future requirements and data from other sources. The current version of the software includes several plugins to manage the five sources introduced in Section~\ref{sec:source}.

First, an analysis of the source's original format is made to determine how to extract the bibliographic metadata and citation data. Once identified, persistent identifiers are processed. Currently, the identifiers managed by OpenCitations, and thus handled within the Source Preprocess step, include DOI, PMID, PMC, VIAF, Wikidata, Wikipedia, ROR, ORCID, Arxiv, JID, ISSN, ISBN, and URL. Each identifier schema handled is managed by particular software, which enables us to:

\begin{enumerate}
  \item normalise identifier values (e.g. converting in lowercase DOIs to standardise their values in the system);
  \item check the correctness of their syntax (e.g. recomputing the check digit of ISSNs to be sure they are structurally sound);
  \item verify their existence using either internal (e.g. in case an identifier has been already validated in the past) or external services (e.g. REST APIs if available).
\end{enumerate}

The software responsible for the aforementioned steps is coupled with another component responsible for storing validated information. Such a storage manager produces two main data collections in the form of tabular information stored in multiple CSV files. On the one hand, it creates CSV files with the \emph{bibliographic metadata} of the processed bibliographic resources involved in citations provided by the input source. On the other hand, it also returns two-column CSV files storing links defined in \emph{citation data}, represented by two entities defined by their persistent identifiers, which may vary depending on the input source. In particular, considering the five sources mentioned in Section~\ref{sec:source}, the Source Preprocess step considers only citation links having the following shape:

\begin{itemize}
    \item DOI-to-DOI citation links for Crossref, DataCite, and JaLC citation data - in this case, only the citation data that specify explicitly in the source data a DOI identifying the cited entities are considered;
    \item PMID-to-PMID citation links for NIH-OCC citation data;
    \item any combination of DOI, PMC, PMID, ArXiv identifier schemas to refer to the citing and cited entities involved in citations for OpenAIRE ScholeXplorer citation data. 
\end{itemize}

The metadata (i.e. table columns) of all the bibliographic metadata and citation data stored in the aforementioned CSV files comply with the OCDM.

\subsubsection{Meta Process}
\label{sec:meta_process}
The main goal of this step is to use the information in the CSV files containing bibliographic metadata obtained in the previous step to get a unique persistent identifier for the citing and cited entity that is internal to the OpenCitations collections. Thus, this identifier can uniquely identify bibliographic resources in the citations gathered from the five sources, independently of the other external persistent identifiers they have associated (DOI, PMID, etc.). 

This identifier is obtained by running the ingestion workflow detailed for the creation of OpenCitations Meta \cite{massari_opencitations_2024}, which is the OpenCitations collection responsible for storing the metadata of all bibliographic resources of all the citations included in the OpenCitations Index. An in-depth analysis of all the features of the OpenCitations Meta ingestion process is described in \cite{massari_opencitations_2024}. In the following text, we summarise only the main operations necessary for creating the OpenCitations Index.

Each bibliographic resource stored in OpenCitations Meta is uniquely identified with a new globally persistent identifier called OpenCitations Meta Identifier (OMID, \url{https://identifiers.org/omid}). In OpenCitations, an OMID -- a simple identifier defined by specifying the prefix ``omid:'' followed by a two-character string defining the type of entity (\emph{br} for bibliographic resources, \emph{ra} for responsible agents, etc.), and finally a ``/'' accompanied by a sequence of digits, such as \texttt{omid:br/06101801781} -- serves as a proxy between various external identifiers used for each bibliographic resource, facilitating deduplication operations if the same bibliographic resource is mentioned in multiple sources. Among the metadata stored in OpenCitations Meta are the publication date, the authors, and the venue of bibliographic resources, which are information necessary for creating part of the metadata of the citations included in the OpenCitations Index introduced in Section~\ref{sec:model}.


While processing the rows in CSV files containing bibliographic metadata extracted from a source, where each row depicts a new bibliographic resource (a journal article, a book chapter, etc.), two main scenarios may occur: 

\begin{enumerate}
    \item the bibliographic resource is already stored in OpenCitations Meta\footnote{The process checks if a bibliographic resource is included in OpenCitations Meta by looking for its external identifiers as specified by the source (DOI, PMID, ISBN, etc.).} and, thus, the related OMID is retrieved;
    \item the bibliographic resource is not available in OpenCitations Meta; thus, a new entity is created, and the related new OMID is generated.
\end{enumerate}
    
The mapping between the original external persistent identifiers available in the CSV files containing the source's citation data and the related OMIDs is tracked in an internal NoSQL database, implemented using Redis (\url{https://redis.io/}) as a Database Management System. These associations are crucial for addressing the last step of the OpenCitations Index ingestion workflow, introduced in the following subsection.

\subsubsection{Index Process}
This phase aims to ingest the citation data gathered from the original source into the OpenCitations Index. The OpenCitations Index software manages the operations conducted at this stage \cite{oc_index_2024}.

This step's input is the CSV file set containing the citation data generated in the Source Preprocess step. Each citation is represented by two entities in these files, which are defined in a two-column tabular format. The citations stored in these tables are links between two potentially persistent identifiers of any type (e.g. DOI-DOI, PMID-PMID, PMC-PMID, etc.) depending on the specific source from which we have gathered them.

To generate citation data to be ingested into the OpenCitations Index, a conversion of the citation links into an OMID-to-OMID format is needed. This operation leverages the content of the Redis database produced in the previous step, i.e. the Meta Process, which includes the mapping between the original persistent identifiers used in the source and the related OMIDs defined as a result of the previous step. For instance, if a citation link extracted from a source was defined using DOIs as persistent identifiers for both the citing and cited entities, we first query Redis to look for the corresponding OMID of each DOI. Then, the input DOI-to-DOI citation link is converted into an OMID-to-OMID one.

At the end of this step, two main outputs are generated. On the one hand, we provide full dumps of citation data compliant with the data model summarised in Figure~\ref{ocdm_citations} we created from the source information. The dumps are provided in RDF (N-triple)\cite{prudhommeaux_rdf_2014}, CSV and SCHOLIX formats, and they include the new citations added to the OpenCitations Index with their corresponding metadata (as introduced in Section~\ref{sec:model}), as shown in Figure~\ref{rdf_cit_1}. On the other hand, we summarise only the OMID-to-OMID citation links in RDF format, with no other metadata specified, which are then uploaded on the OpenCitations Index graph database (i.e. an RDF triplestore) created using QLever (\url{https://github.com/ad-freiburg/qlever}), an engine that can efficiently index and query very large knowledge graphs.

\begin{figure}[hbtp]
    \label{rdf_cit_1}
    \begin{lstlisting}[language=Mathematica,frame=single]
@prefix br: <https://w3id.org/oc/meta/br/> .
@prefix ci: <https://w3id.org/oc/index/ci/> .
@prefix cito: <http://purl.org/spar/cito/> .
@prefix rdf: <http://www.w3.org/1999/02/22-rdf-syntax-ns#> .
@prefix xsd: <http://www.w3.org/2001/XMLSchema#> .

ci:06101801781-06180334099
    rdf:type cito:Citation ;
    cito:hasCitingEntity br:06101801781 ;
    cito:hasCitedEntity br:06180334099 ;
    cito:hasCitationCreationDate "2021-03-10"^^xsd:date ;
    cito:hasCitationTimeSpan "P6Y0M1D"^^xsd:duration .
    \end{lstlisting}
    \caption{The RDF Turtle format of a citation stored and modelled according to the OCDM.}
\end{figure}

Following this organisation, every citation is uniquely identified by a persistent identifier, i.e. the \emph{Open Citation Identifier} (OCI, \url{https://identifiers.org/oci}). The OCI has been created and maintained by OpenCitations \cite{Peroni2019}, and has a simple structure: the lower-case letters ``oci'' followed by a colon, followed by two sequences of digits separated by a dash. For example, \texttt{oci:06101801781-06180334099} is a valid OCI for citations defined within the OpenCitations Index, where the first sequence of digits represents the identifier of the bibliographic resource (\emph{br}) having numerical identifier ``06101801781'' (i.e. the OMID \texttt{omid:br/06101801781}), while the second sequence is the bibliographic resource having numerical identifier ``06180334099'' (i.e. the OMID \texttt{omid:br/06180334099}). The two sequences of digits in an OCI (e.g. ``06101801781-06180334099'') define the unique identifier for citations in the OpenCitations Index according to the OCDM (i.e. the entity \url{https://w3id.org/oc/index/ci/06101801781-06180334099}, or \texttt{ci:06101801781-06180334099} in abbreviated form, as shown in Figure~\ref{rdf_cit_1}).

\subsection{Provenance, change tracking, and dataset metadata}
In addition to citation data stored in the OpenCitations Index, the ingestion process developed is also responsible for providing details about the provenance and versioning of each citation. In particular, following the specification of the OCDM, which reuses terms from the Provenance Ontology (PROV-O, \url{https://www.w3.org/ns/prov-o}) \cite{lebo_prov-o:_2013}, every citation is defined by one or more provenance snapshots (i.e. instances of \texttt{prov:Entity} that capture the status of such entity (referred via the property \texttt{prov:specializationOf}) at a given time. Indeed, each snapshot records validity and invalidity dates, responsible agents for either the creation or the modification of the metadata, primary data sources, and a SPARQL query summarising changes with respect to any prior snapshot, which is implemented via an appropriate property defined in the OCDM (i.e. \texttt{oco:hasUpdateQuery}). In addition, the \textit{prov:atLocation} is used to keep track of the internal collection (one for each data source used) where citation are organised.

\begin{figure}
\includegraphics[width=\textwidth]{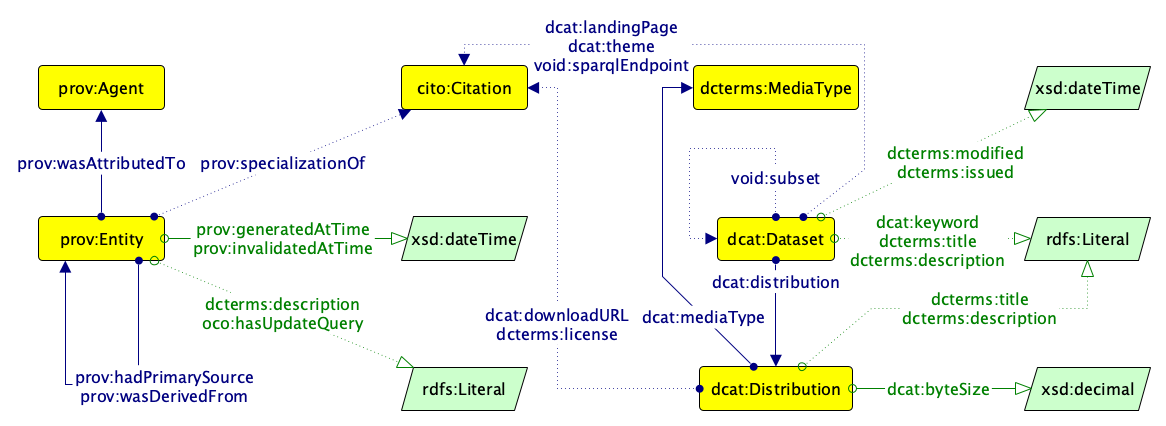}
\caption{The section of the OCDM that is dedicated to provenance, change tracking information, and dataset metadata.}\label{ocdm_prov}
\end{figure}

Finally, we also include a metadata description of the entire dataset defined by the OpenCitations Index, keeping track of when it is modified, and specifying each produced dump distribution created and published alongside the collection. For describing these data, we reuse two additional ontologies, i.e. the Vocabulary of Interlinked Datasets (VoID, \url{http://rdfs.org/ns/void}) \cite{alexander_describing_2009} and the Data Catalog Vocabulary (DCAT, \url{http://www.w3.org/ns/dcat}) \cite{albertoni_data_2024}. In particular, a dataset (\texttt{dcat:Dataset}) containing information about the citations is described with cataloguing information (e.g. title, description, publication and change dates, subjects, webpage, SPARQL endpoint) and distribution information (\texttt{dcat:Distribution}) which also includes the specification of licenses, dumps, media types, and data volumes.

\subsection{Dataset}

As of July 2024, the OpenCitations Index stores 2.01 billion unique citation links, between over 91,38 million bibliographic resources -- in detail, 72.8 million unique citing entities and over 74.2 million unique cited entities. All the data are downloadable in dumps in several formats (CSV, N-triple, and Scholix, downloadable from \url{https://opencitations.net}), and each citation can be systematically accessed through its HTTP URLs. 
The citation data are accompanied by provenance and change tracking information (in CSV and N-Triples).

\begin{figure}
\includegraphics[width=\textwidth]{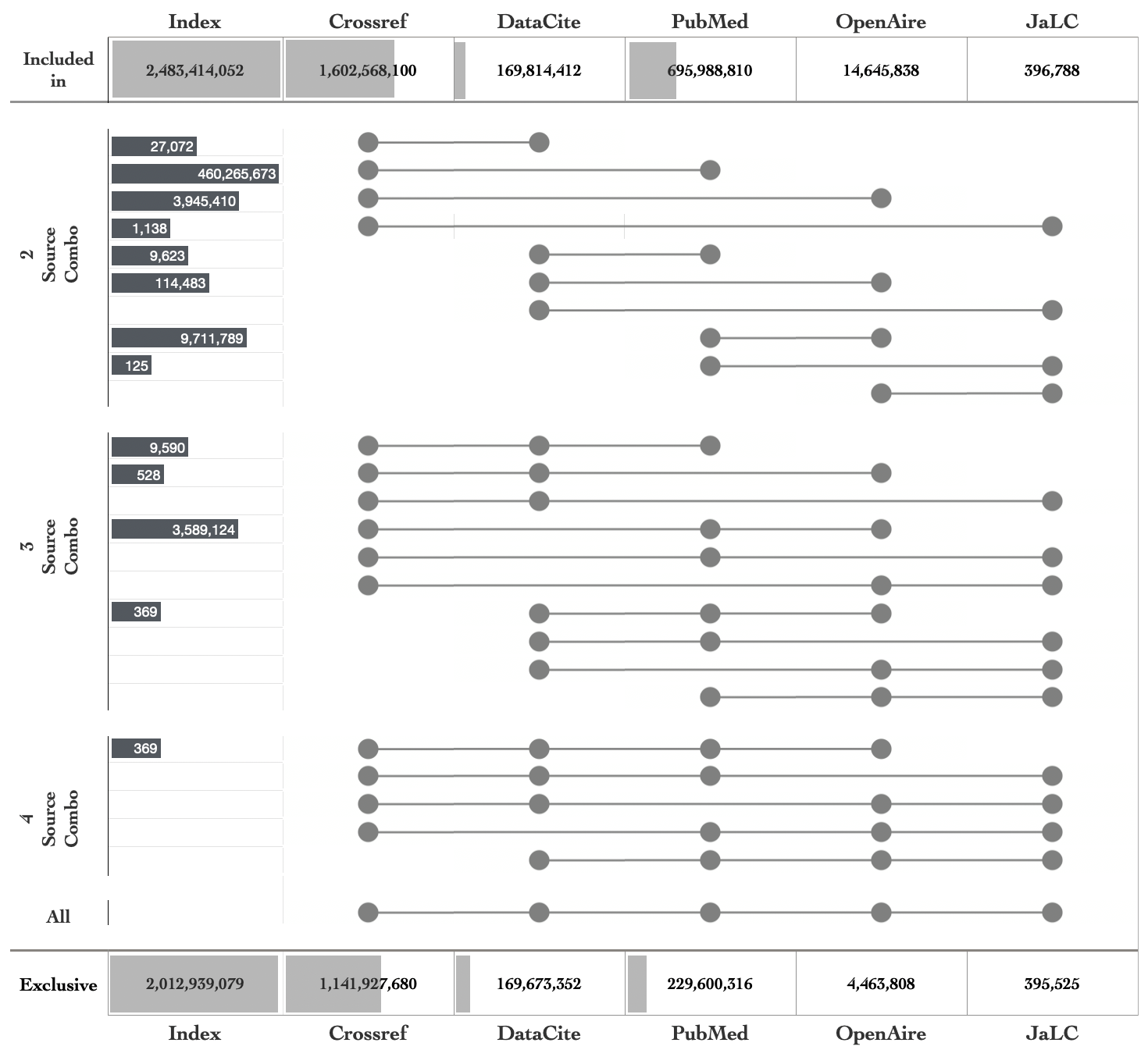}
\caption{An overview of the number of citations stored in the OpenCitations Index as of July 2024. Each column represents a different collection (data source), while each row represent the citations that are shared between the collections (overlapping citations), such that the combined collections are graphically marked with a dot. The last row (i.e. Exclusive) show the unique contribution of each collection to the OpenCitations Index.}
\end{figure}
\label{covarage}

Figure~\ref{covarage} provides an overview of the contribution of each source. The most of the citation information comes from Crossref, which provides 1,602,568,100 of the 2,012,939,079 total citations. Its unique contribution (i.e. citation links that are contained only in Crossref) is of 1,141,927,680 citations. The NIH-OCC is the second source by overall contribution with 695,988,810 citation links, although its unique contribution is just over one-third of the data provided (i.e. 229,600,316). Datacite provides 169,814,412 citations, and the sharp majority (i.e. 169,673,352 citations) were not present in any of the other sources. Similarly, JaLC provides 396,788 citations, but only a small portion of these were also present in other sources (specifically, 1,138 in Crossref and 125 in NIH-OCC). In contrast, of the 14,645,838 references provided by OpenAIRE, 4,463,808 are the exclusive contribution of the source. It is worth mentioning that there are no citations in common between all the collections (the raw "All").

All data are released under a CC0 waiver, to maximise the unrestricted use, transformation, and integration of the OpenCitations Index data into other systems and workflows. Further, this approach enhances discoverability, interoperability, and reusability of citation data, thereby contributing to scholarly knowledge at large.

\subsection{Resources and services}
To maximise interoperability and reuse of the exposed information, OpenCitations offers a comprehensive suite of tools designed to facilitate programmatic access to its citation data, meeting the needs of a wide range of users from Semantic Web experts to web developers and researchers. In particular, it is possible to access citation data included in the OpenCitations Index using its SPARQL endpoint (\url{https://w3id.org/oc/index/sparql}, that support queries through the SPARQL Protocol And RDF Query Language (or simply SPARQL) \cite{harris_sparql_2013}. In addition, we also make available a REST API (\url{https://w3id.org/oc/index/api/v2}), implemented using RAMOSE \cite{daquino_creating_2022}, to offer a straightforward way to access data via HTTP REST calls, simplifying the integration of OpenCitations data into diverse applications. Additionally, for searching, querying, and browsing data, we either adopt or develop Web-based applications that include  YASGUI (SPARQL Web GUI) \cite{rietveld_yasgui_2016}, OSCAR \cite{heibi_enabling_2019} and LUCINDA (\url{https://github.com/opencitations/lucinda}).

\section{Usage}\label{sec:usage}
In this section, we delve into an analysis of the utilisation of OpenCitations Index. In particular, we focus on two distinct facets: a quantitative analysis, providing a statistical overview of the available services and resources, and the community uptake, which examines the impact of OpenCitations Index on the broader community.

\subsection{Quantitative analysis}

\begin{figure}
\includegraphics[width=\textwidth]{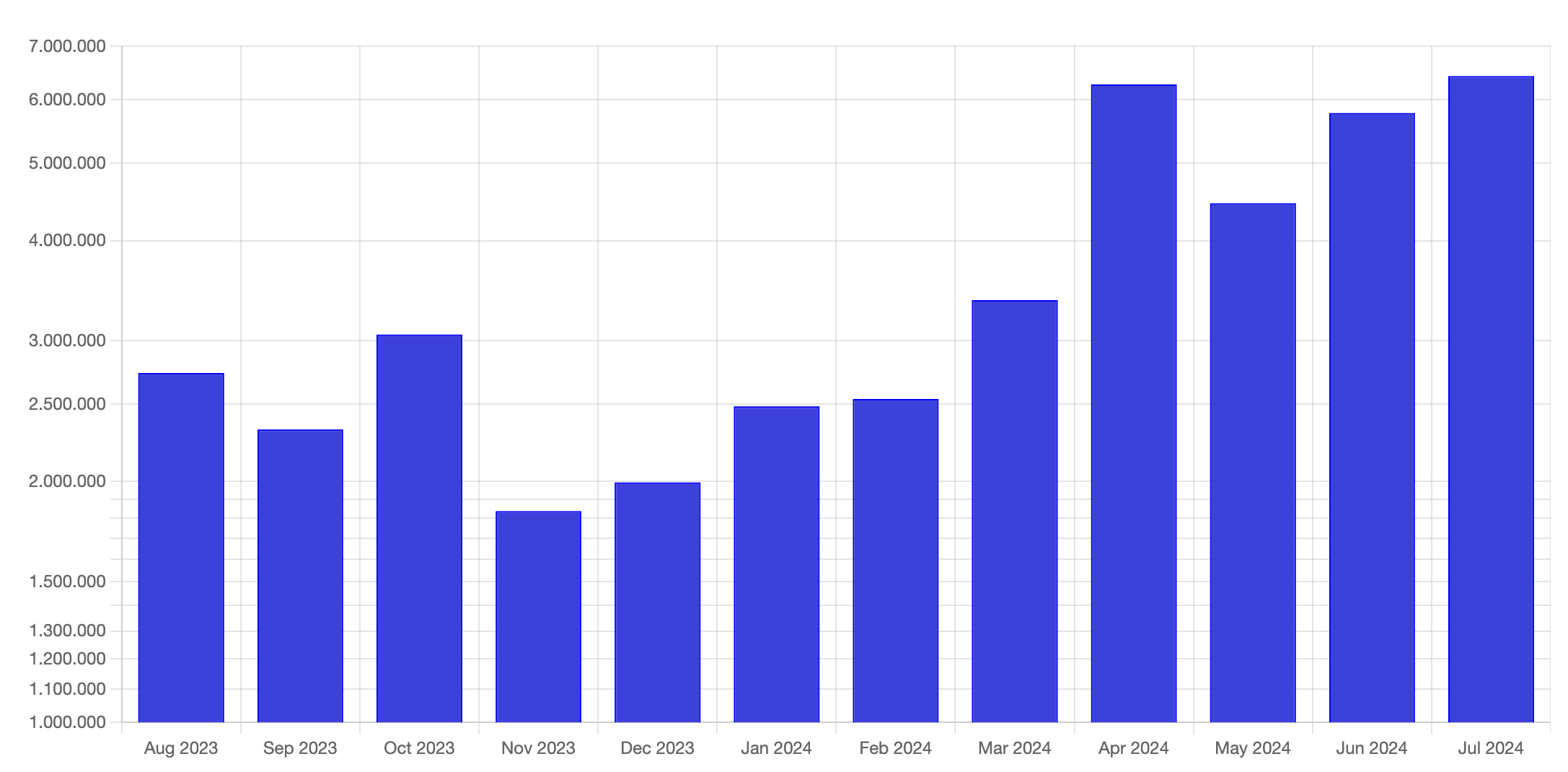}
\caption{The total number of accesses to the OpenCitations Index API service between August 2023 and July 2024}
\label{tot_api_index}
\end{figure}

Figure \ref{tot_api_index} depicts the number of accesses recorded between August 2023 and July 2024 (inclusive) to the REST API of the OpenCitations Index. For these statistics, accesses made by automated agents and bots have been excluded to the best of our knowledge. July marks the end of this reporting period and also records the highest number of accesses, exceeding 6.4 million API calls.

\begin{figure}
\includegraphics[width=\textwidth]{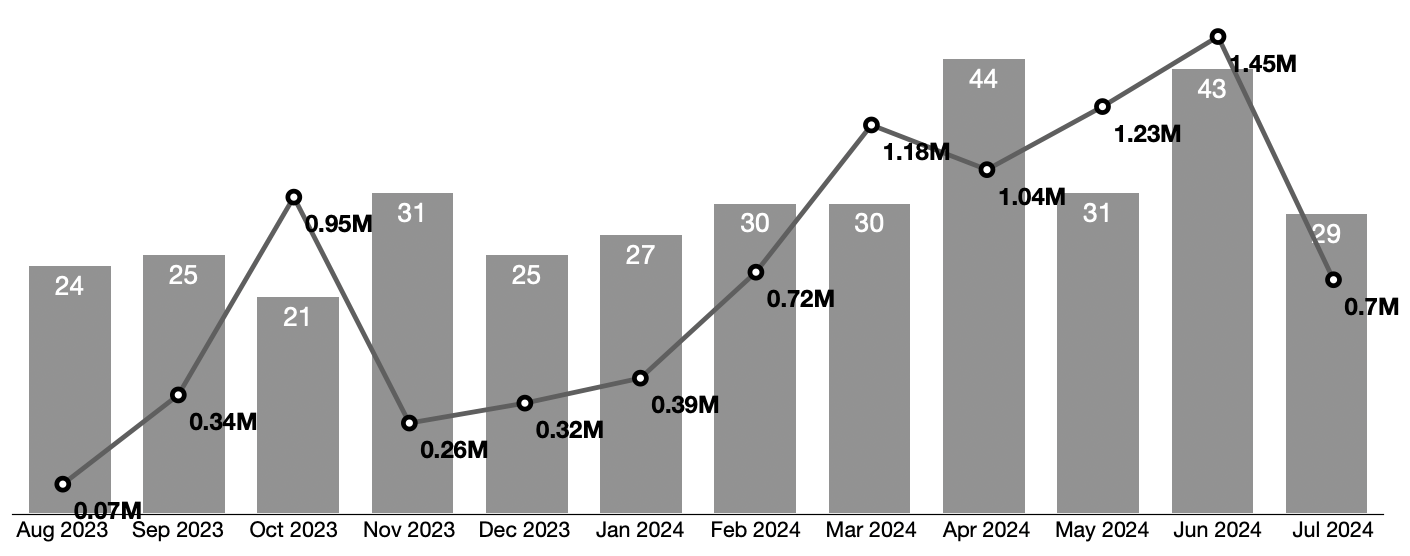}
\caption{A monthly projection of the number of authenticated users using OpenCitations services (represented by bars) alongside their corresponding total number of calls made (represented by a line)}
\label{tokens}
\end{figure}

The OpenCitations policy grants all users worldwide free access to its services without requiring registration or sign-up. To enable us to track the effective number of users using and accessing the services, creating an anonymous access token is possible for users who desire to help us with tracking the number of users (e.g. applications) adopting our services. Figure~\ref{tokens} displays, using bars, unique users (authenticated via tokens) and the number of monthly queries, represented by the line above. The average number of authenticated users that use the API services of OpenCitations per month, considering the last year, is 30. We also reported 80 active users, i.e., users that used the REST API at least once in the last year. Of these, 13 remained active throughout the entire year period, making at least one call each month. 

\begin{figure}
\includegraphics[width=\textwidth]{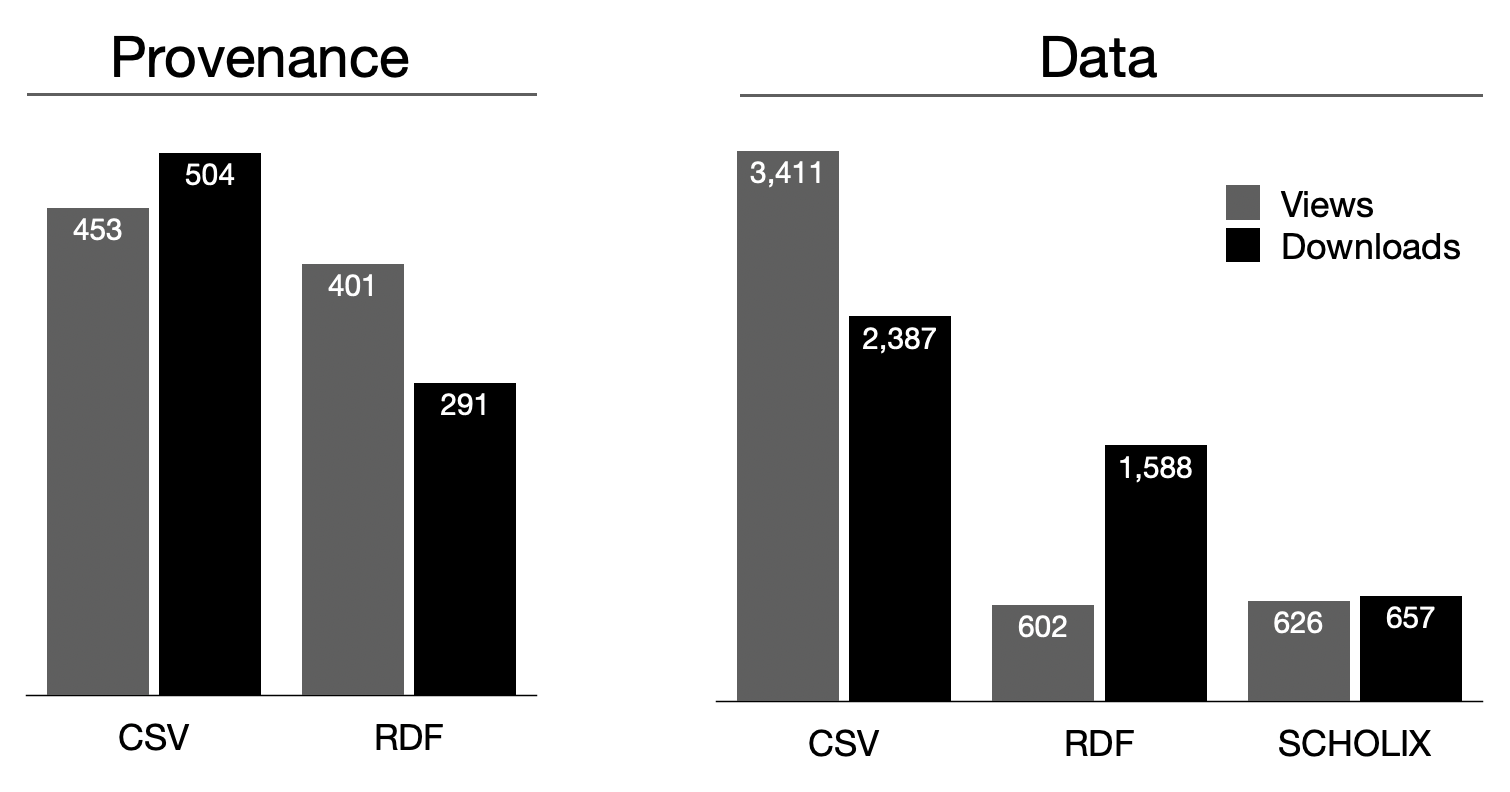}
\caption{The total number of downloads and views of the OpenCitations Index data (i.e. citations) and provenance datasets in all the available formats (CSV, RDF, and SCHOLIX for the citation dataset and CSV and RDF for the provenance dataset), retrieved from FigShare, as of July 26, 2024.}
\label{dataset_and_prov}
\end{figure}

Information about views and downloads of the entire OpenCitations Index dataset (citations and provenance) have been gathered from Figshare. On the one hand, from 25 October 2023, the date of the first publication of the new OpenCitations Index on FigShare, to 26 July 2024, the dataset of citations in CSV format was downloaded 2,387 times and viewed 3,411 times. In contrast, the N-Triples (RDF) version counts 1,588 downloads and 602 views. Finally, the dataset was downloaded 657 times in SCHOLIX format and viewed 626. On the other hand, the provenance dataset has 504 downloads and 453 views for the CSV format, and 291 downloads and 401 views for the RDF format. These statistics are represented in the charts of Figure \ref{dataset_and_prov}. The downloads are, in some cases, higher than the views because the items can also be downloaded using the API service, thus not raising the views counter.

\subsection{Community uptake}
In addition to the quantitative statistics presented in the previous subsection, the citation data provided by OpenCitations are used by several independent international initiatives and projects. 



In June 2023, the OpenAIRE-Nexus H2020 project (\url{https://www.openaire.eu/openaire-nexus-project} came to an end, with the release of the new version of the OpenAIRE Graph, which included all citations available in OpenCitations. In another Horizon Europe project, i.e. GraspOS (\url{https://graspos.eu/}), OpenCitations is involved as one of the data asset sources and provides raw and structured citation data (included in the OpenCitations Index) that can be used to directly support Open Science-aware responsible research assessment processes, e.g. by developing indicators and collecting evidence that can be of value in these processes. 

Other projects and initiatives have started to use OpenCitations Index data as valuable sources. For instance, B!SON (\url{https://service.tib.eu/bison/}) is a web-based journal recommendation system that integrates multiple data sources, including the Directory of Open Access Journals (DOAJ, \url{https://doaj.org}) and OpenCitations, to aid authors and library publication support services by recommending suitable open-access journals \cite{entrup_comparing_2023}. 

PURE Suggest (\url{https://fabian-beck.github.io/pure-suggest/}) is a scientific literature search tool. It recommends scholarly articles based on citations (gathered from OpenCitations) and references from initial seed papers \cite{beck_visually_2022}. 

The Open Access Helper (OAHelper, \url{https://www.oahelper.org/}), instead, is a browser extension and an app designed to assist users in finding legal, open-access versions of scientific articles. To accomplish this task and provide additional metadata and citation information about articles, the OAHelper uses, among the others, OpenCitations Index API to retrieve relevant citation data.

In addition to the aforementioned projects, also some institutional repositories have started to use OpenCitations data. ORBi (\url{https://projects.tib.eu/optimeta/en/}) is the institutional repository of the University of Liège. ORBi aims to collect, preserve, and provide free access to the publication output of the university's researchers, thus enhancing the visibility and impact of the research conducted at the University of Liège. In ORBi, citations counts are gathered from different indexes, including OpenCitations. 

CHERRY (\url{https://cherry.chem.bg.ac.rs/}) is a joint digital repository of all departments in the Faculty of Chemistry, University of Belgrade. It provides open access to this institution's publications and other research outputs. It allows browsing and searching for authors and funding information and displays Altmetric scores and Dimensions, Scopus, OpenCitations, and Web of Science citation counts. 

StabiKat (\url{https://stabikat.de/}) is an index of more than 190 million articles, books, e-books, digital items, and more from the collections of the Staatsbibliothek zu Berlin. These items are accompanied by metrics, such as citations, retrieved from OpenCitations. 

\section{Conclusions}\label{sec:conclusions}
In this paper, we have introduced the OpenCitations Index -- a database of open citation data curated by OpenCitations. The Index is built upon a collection of citation data gathered from various sources. This dataset offers added value compared to previous citation datasets managed by OpenCitations, as it represents a strategic change in the data ingestion approach, moving into having one unique and deduplicated storage for citations. This is achieved by excluding duplicate citation links that may arise from multiple sources referring to the same citations between the same bibliographic resources. We have introduced the new workflow for constructing such a dataset, which relies on another curated dataset, OpenCitations Meta, and the new persistent identifier maintained by OpenCitations, i.e. the OpenCitations Meta Identifier (OMID).

Following the positive trend and the usage of the community, the plan is to improve the quality of OpenCitations Index data further. We are working on Web-based interfaces for provenance-aware human curation of citation data. In particular, we have released the first version of HERITRACE \cite{massari_heritrace_2024}, that is a novel semantic data management system which provide customisable Web form for modifying RDF data keeping track of provenance information and tracking data changes. We will use HERITRACE to create a specific Web-interface for OpenCitations data curation, which will enable us to add and modify (i.e. correct) citation data included in all OpenCitations collections, including the OpenCitations index. Thanks to this interface, we will improve the quality of OpenCitations citation data by involving human curators in the loop. In addition, we aim to improve the data quality by developing and adopting computational-aware author's disambiguation approaches, combining machine learning techniques with network analysis.

\begin{credits}
\subsubsection{\ackname} This work has been partially funded by the European Union’s Horizon Europe framework programme under grant agreement No 101095129 (GraspOS Project).

\subsubsection{Author Contributions}
\begin{itemize}
    \item \textbf{Ivan Heibi:} Conceptualization, Data curation, Formal Analysis, Methodology, Software, Validation, Visualization, Writing – original draft, Writing – review \& editing
    \item \textbf{Arianna Moretti:} Conceptualization, Data curation, Formal Analysis, Methodology, Software, Validation, Visualization, Writing – original draft, Writing – review \& editing
    \item \textbf{Marta Soricetti:} Methodology, Software, Validation, Writing – original draft, Writing – review \& editing
    \item \textbf{Silvio Peroni:} Funding acquisition, Project administration, Resources, Supervision, Writing – original draft, Writing – review \& editing

\end{itemize}

\subsubsection{Data Accessibility Statement.}
All the Index distributions are deposited on Figshare under a CC0 license and can be downloaded in different formats, alongside their provenance information.

Citation Dataset:
\begin{itemize}
    \item \textbf{OpenCitations Index CSV dataset of all the citation data}\footnote{https://doi.org/10.6084/m9.figshare.24356626}.
\end{itemize}
\begin{itemize}
    \item \textbf{OpenCitations Index N-Triples dataset of all the citation data}\footnote{https://doi.org/10.6084/m9.figshare.24369136}.
\end{itemize}
\begin{itemize}
    \item \textbf{OpenCitations Index Scholix dataset of all the citation data}\footnote{https://doi.org/10.6084/m9.figshare.24416749}.
\end{itemize}
Provenance Dataset:
\begin{itemize}
    \item \textbf{OpenCitations Index CSV dataset of the provenance information of all the citation data}\footnote{https://doi.org/10.6084/m9.figshare.24417733}.
\end{itemize}
\begin{itemize}
    \item \textbf{OpenCitations Index N-Triples dataset of the provenance information of all the citation data}\footnote{https://doi.org/10.6084/m9.figshare.24417736}.
\end{itemize}

\subsubsection{\discintname}
The authors have no competing interests to declare that are relevant to the content of this article.
\end{credits}

%
%

\bibliographystyle{apacite}
\bibliography{main}
\end{document}